\documentclass[reprint,superscriptaddress,amsmath,amssymb,prb]{revtex4-1}

\usepackage{bm}
\usepackage{graphicx}

\makeatletter
\AtBeginDocument{\let\LS@rot\@undefined}
\makeatother

\begin{document}

\title{Breaking the fundamental energy dissipation limit in
       ferroelectric-dielectric capacitors}
\date{\today}

\author{Justin C. Wong}
\affiliation{Department of Electrical Engineering and Computer Sciences,
             University of California, Berkeley, California 94720, USA.}
\author{Sayeef Salahuddin}
\email[Correspondence and requests for materials should be addressed to Sayeef
       Salahuddin: ]{sayeef@berkeley.edu}
\affiliation{Department of Electrical Engineering and Computer Sciences,
             University of California, Berkeley, California 94720, USA.}
\affiliation{Department of Materials Science and Engineering, University of
             California, Berkeley, California 94720, USA.}

\begin{abstract}
  Half of the energy is always lost when charging a
  capacitor\cite{Salahuddin2007}. Even in the limit of vanishing resistance,
  half of the charging energy is still lost---to radiation instead of heat.
  While this fraction can technically be reduced by charging adiabatically, it
  otherwise places a fundamental limit on the charging efficiency of a
  capacitor. Here we show that this $1 / 2$ limit can be broken by coupling a
  ferroelectric to the capacitor dielectric. Maxwell's equations are solved for
  the coupled system to analyze energy flow from the perspective of Poynting's
  theorem and show that (1) total energy dissipation is reduced below the
  fundamental limit during charging and discharging; (2) energy is saved by
  ``recycling'' the energy already stored in the ferroelectric phase transition;
  and (3) this phase transition energy is directly transferred between the
  ferroelectric and dielectric during charging and discharging. These results
  demystify recent works\cite{Rusu2010,IslamKhan2011,Lee2013,Appleby2014,
  Khan2014,Lee2015,Khan2016,Zubko2016} on low energy negative capacitance
  devices as well as lay the foundation for improving fundamental energy
  efficiency in all devices that rely on energy storage in electric fields.
\end{abstract}

\maketitle

Ferroelectrics have recently been shown to exhibit a negative capacitance effect
\cite{Salahuddin2008,IslamKhan2011,Zubko2016} when placed in a series
combination with a dielectric film. Under appropriate conditions, the dielectric
leads to a strong depolarization field that forces the ferroelectric into its
normally unstable near-zero polarization states. These results have garnered
considerable interest in ferroelectric negative capacitance due to its potential
to reduce power consumption below thermodynamic limits in electronic
devices\cite{Salahuddin2008}. In field-effect transistors, for example, negative
capacitance has been proposed as a solution to end the ``Boltzmann tyranny'' on
subthreshold swing\cite{Salahuddin2008,zhirnov2008nanoelectronics,theis2010s}.
However, negative capacitance has traditionally been understood to require a
source (e.g. a battery) that supplies extra energy\cite{Jonscher1986}. This begs
the following question: if extra energy must be supplied by some source, then
where is the energy coming from in ferroelectric negative capacitance and is any
energy truly saved?

\begin{figure*}
  \includegraphics[width=\linewidth]{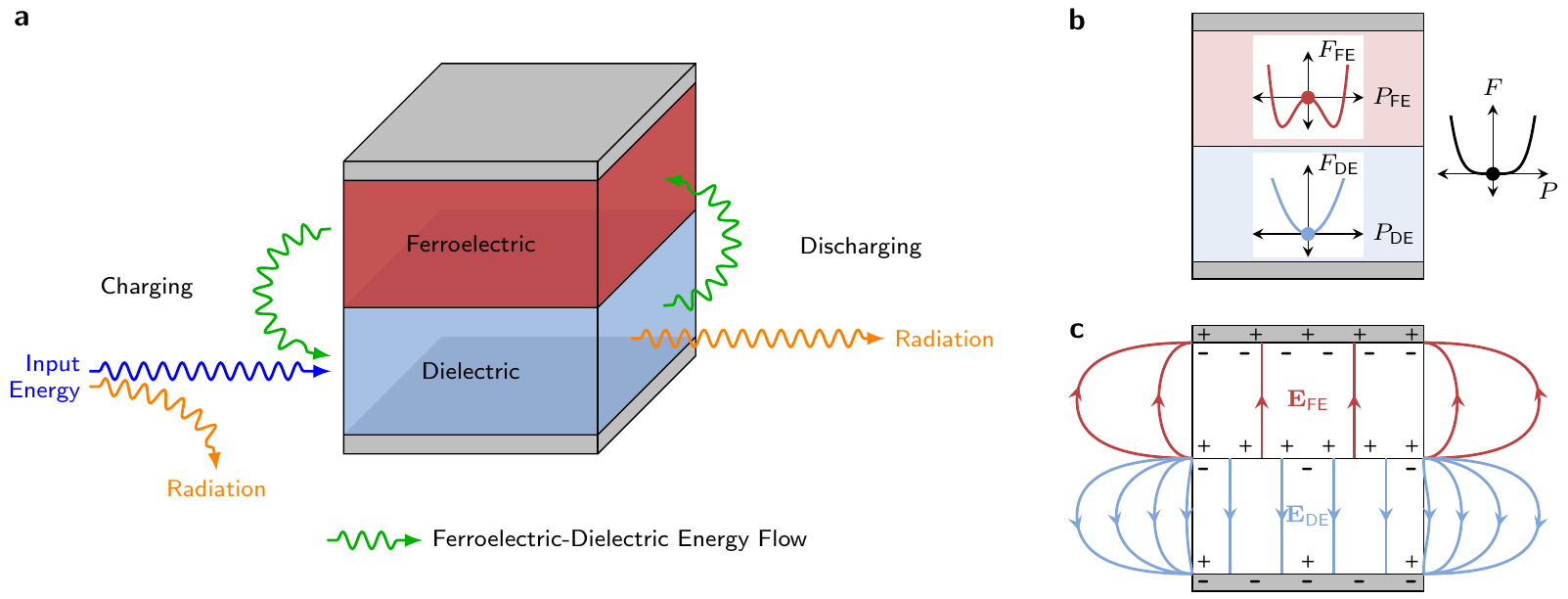}
  \caption{Paths of energy flow during charging and discharging of a capacitor
           with a ferroelectric. (a) Schematic of energy flow paths during
           charging (left) and discharging (right) of a ferroelectric-dielectric
           capacitor. New paths of energy flow emerge between the ferroelectric
           and dielectric during charging and discharging. These paths are not
           present in conventional dielectric capacitors. (b) Energy landscapes
           in a ferroelectric-dielectric capacitor. If the dielectric is
           sufficiently thick, then its energy landscape $F_{\text{DE}}$ will
           dominate the energy landscapes $F_{\text{FE}}$ and $F$ of the
           ferroelectric and overall system respectively. The dielectric
           polarization $P_{\text{DE}}$ then forces the ferroelectric near its
           phase transition at zero polarization ($P_{\text{FE}} = 0$) via a
           strong depolarization field. This puts the ferroelectric into a
           higher energy state in which energy can be extracted from the phase
           transition. (c) Schematic of the total electric fields in a
           ferroelectric-dielectric capacitor. When an external electric field
           is applied, the ferroelectric and dielectric both polarize by
           different amounts, resulting in a depolarization field. Since the
           ferroelectric is stabilized in a higher energy state near zero
           polarization, it releases energy when polarized. This extra energy
           contributes towards further strengthening the depolarization field,
           which subsequently further polarizes and charges the dielectric. The
           resultant electric fields $\mathbf{E}_{\text{FE}}$ and
           $\mathbf{E}_{\text{DE}}$ in the ferroelectric and dielectric
           respectively end up pointing in opposite directions.}
  \label{fig:schematic}
\end{figure*}

To answer this question, we considered energy flow during charging and
discharging of a ferroelectric-dielectric capacitor. We solved Maxwell's
equations for the coupled system and used Poynting's theorem to show how energy
flows. We performed our analysis in the limit of zero resistance in order to
understand the fundamental charging and discharging energy costs. The results
are shown schematically in Fig. \ref{fig:schematic}a.
During charging, input energy flows from an energy source to the dielectric, and
a fraction of that energy is dissipated. This dissipation is dominated by
electromagnetic radiation in the limit of zero resistance. Notice that there is
an additional path of energy flow from the ferroelectric to the dielectric that
is not present during charging in conventional capacitors. This implies that the
ferroelectric is supplying extra energy to the dielectric.  During discharging,
the dielectric acts as the energy source and normally dissipates all of its
energy when in a conventional capacitor. However, there is a new path of energy
flow that allows the dielectric to transfer a fraction of its energy back to the
ferroelectric. Thus, we see schematically how energy may be internally recycled
in the coupled ferroelectric-dielectric system. However, it is still unclear
where the extra energy comes from and how it transfers between the ferroelectric
and dielectric.

The origin of this extra energy can be understood from a thermodynamic
perspective as shown in Fig. \ref{fig:schematic}b. Due to their phase
transition, ferroelectrics possess a higher energy, zero polarization state in
their energy landscape. This is in contrast to dielectrics, which have a minimum
in their energy landscape at zero polarization. Consequently, coupling a
ferroelectric to a dielectric results in a large divergence in polarization at
the interface. This establishes a strong depolarization field that stabilizes
the ferroelectric near its unstable zero polarization state. The net effect is
an electrically-induced transition towards a phase of higher crystal symmetry
and can be thought of as an effective shift in the phase transition temperature
\cite{Salahuddin2008,Zubko2016}. This electrical influence is in conflict with
the natural temperature-induced transition towards lower crystal symmetry. Thus,
we can electrically extract energy from the phase transition by modulating this
conflict with an applied electric field. The extracted energy is then
transferred between the ferroelectric and dielectric via propagation of the
depolarization field as shown schematically in Fig. \ref{fig:schematic}c. Notice
that the electric field points in opposite directions from the
ferroelectric-dielectric interface due to the negative permittivity of the
ferroelectric near its phase transition. It is worth noting that this result was
directly obtained from our calculations without any consideration a priori of
negative electric susceptibilities or capacitances.

For our quantitative analysis, we modelled the coupled ferroelectric-dielectric
system using the electric Gibbs free energy (which we will refer to as simply
free energy for the remainder of this letter):
\begin{equation} \label{eq:free_energy}
  G = \int \left( f - \mathbf{E} \cdot \mathbf{P} \right) \, d^{3}r
\end{equation}
$f$ is the Helmholtz free energy density as a function of temperature and
polarization $\mathbf{P}$; and $\mathbf{E}$ is electric field. Note that $f$,
$\mathbf{E}$, and $\mathbf{P}$ all vary with position $\mathbf{r}$, and the
functional form of $f$ depends on the material energy landscape. The dielectric
was modelled as a linear dielectric, and its electric susceptibility and
thickness were normalized as a single tuning parameter. The ferroelectric energy
landscape was modelled after Pb(Zr$_{0.52}$Ti$_{0.48}$)O$_{3}$ using
Landau-Devonshire phenemonological parameters\cite{Rabe2007}. We could have used
Ginzburg-Landau theory to take into account slow spatial variations in the
polarization. However, such fine details would simply add finer spatial
variations to our calculated energy flow; the overall flow would remain the same
as long as the ferroelectric was still locally stabilized near its zero
polarization state. We also assumed a one-dimensional order parameter since
polarization is expected to lie primarily along the capacitor axis. Finally, we
solved for the stationary states of the coupled ferroelectric-dielectric system
by minimizing the free energy with respect to small polarization fluctuations
$\delta \mathbf{P}$ under constant electric field and isothermal conditions.

\begin{figure*}
  \includegraphics[width=\textwidth]{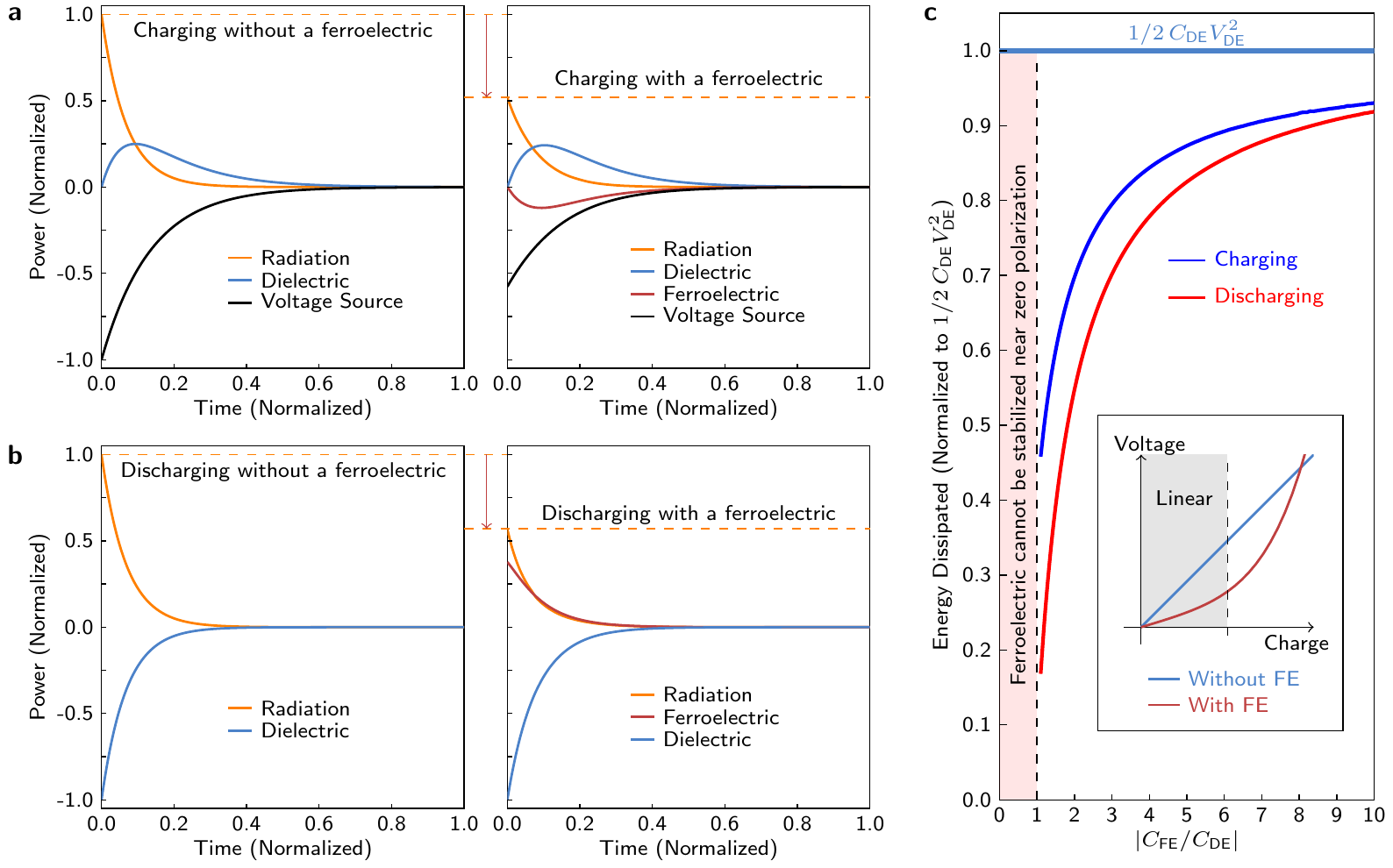}
  \caption{Power comparison during charging and discharging of a capacitor
           without/with a ferroelectric. Negative power corresponds to power
           supplied. (a) Power versus time during charging of a capacitor
           without a ferroelectric (left) and with a ferroelectric added
           (right). The ferroelectric and dielectric parameters are set such
           that $|C_{\text{FE}} / C_{\text{DE}}| = 2$ (see (c) and Fig.
           \ref{fig:matching} for more information) where $C_{\text{FE}}$ is the
           ferroelectric capacitance, and $C_{\text{DE}}$ is the dielectric
           capacitance. The voltage source supplies less power when the
           ferroelectric is coupled to the dielectric, and the amount of power
           radiated is reduced. The dielectric still receives the same amount of
           energy because the ferroelectric supplies the missing power. (b)
           Power versus time during discharging of the same capacitors from (a)
           (without a ferroelectric, left; and with a ferroelectric, right). The
           dielectric acts as the source during discharging, and a fraction of
           its power is delivered to the ferroelectric instead of completely
           radiating away as in the conventional case. (c) Total energy
           dissipated as a function of the capacitance matching
           $|C_{\text{FE}}/C_{\text{DE}}|$ after charging and discharging. The
           energy is normalized to $1 / 2 \, C_{\text{DE}} V_{\text{DE}}^2$,
           which is the conventional amount of energy dissipated without a
           ferroelectric. The inset shows that the capacitor becomes nonlinear
           when a ferroelectric is added, resulting in charge-dependent energy
           dissipation. The curves shown here were calculated by charging to and
           discharging from the end of the linear region.}
  \label{fig:power}
\end{figure*}

The dynamics of the system were described with Poynting's theorem:
\begin{equation} \label{eq:poynting_theorem}
  \frac{\partial u}{\partial t} + \nabla \cdot \mathbf{S} = -\mathbf{J}
    \cdot \mathbf{E}
\end{equation}
The internal energy density $u$ inside the ferroelectric and dielectric was
determined by solving for the states of the coupled ferroelectric-dielectric
system (as described in the previous paragraph). The form of the Poynting vector
$\mathbf{S}$ was obtained by solving Maxwell's equations using retarded scalar
and vector potentials. For ease of calculation, we considered a simple wire loop
geometry containing a voltage source and the ferroelectric-dielectric capacitor
at diametrically opposite positions. With $u$ and $\mathbf{S}$, it is possible
to numerically solve the differential equation (\ref{eq:poynting_theorem}) and
compute the power supplied and consumed over time as shown in Fig.
\ref{fig:power}a-b.
During charging (Fig. \ref{fig:power}a), less power is supplied by the energy
source when a ferroelectric is coupled to the dielectric, and less power is lost
to radiation. However, the dielectric still receives the same amount of energy,
and we find that the ferroelectric supplies the missing part using the energy
stored in its phase transition. During discharging (Fig. \ref{fig:power}b), the
dielectric radiates energy, but the ferroelectric appears to ``capture'' a
fraction of it back to replenish its phase transition energy when coupled to the
dielectric. The total energy dissipated is shown in Fig. \ref{fig:power}c as a
function of the ``capacitance matching'' between the ferroelectric and
dielectric. We find that the amount of energy dissipated is reduced below the
conventional $1 / 2 \, C V^{2}$ during charging and discharging, and it is
minimized when the ferroelectric and dielectric capacitances are equal (i.e.
$|C_{\text{FE}} / C_{\text{DE}}| \to 1$). It should be noted that the capacitor
becomes nonlinear when a ferroelectric is added (inset of Fig.
\ref{fig:power}c), resulting in charge-dependent energy dissipation. We charged
to the end of the linear region and then discharged completely to calculate the
curves in Fig. \ref{fig:power}c.

\begin{figure*}
  \includegraphics[width=\textwidth]{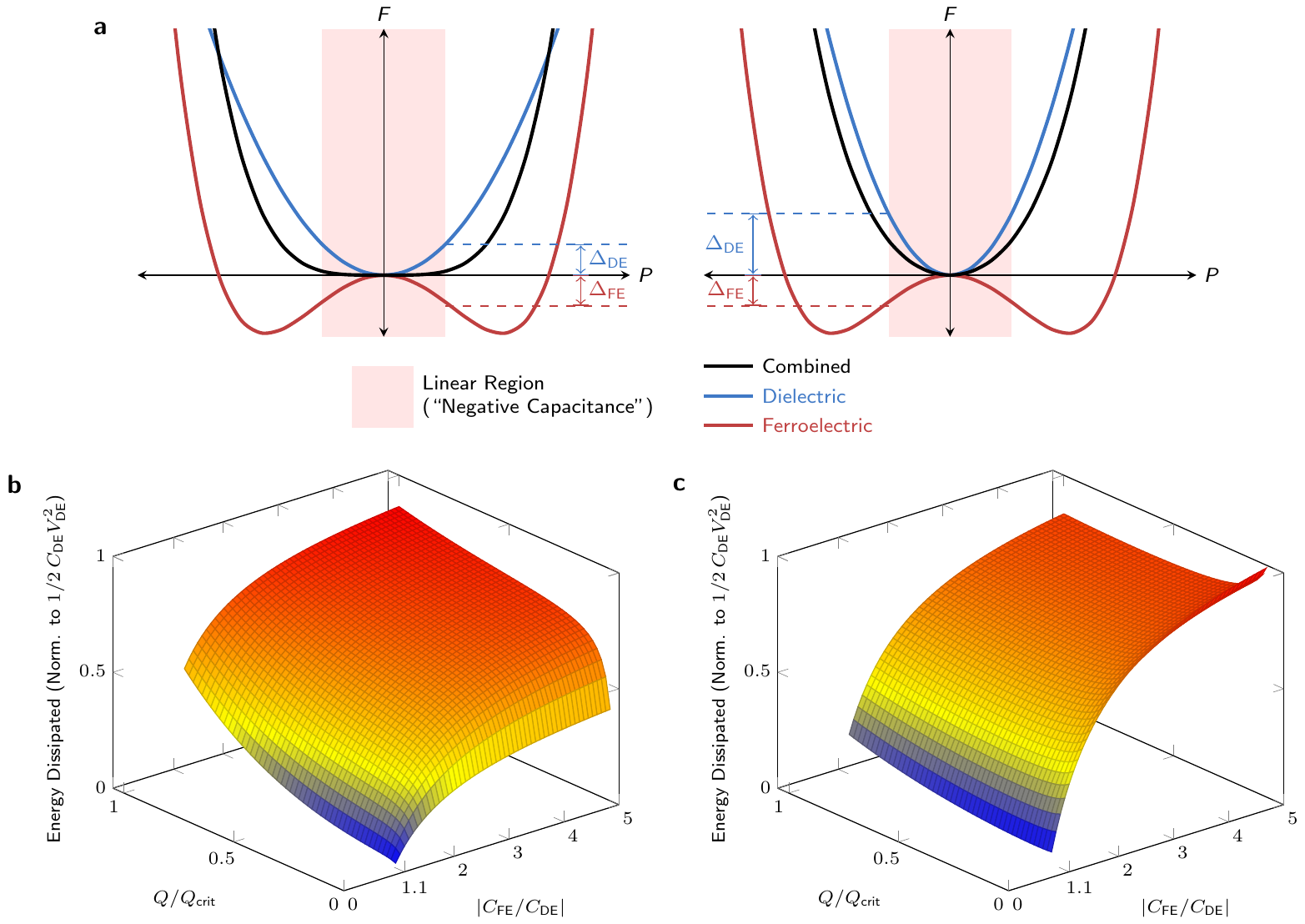}
  \caption{Energy balancing and capacitance matching. (a) Energy landscapes
           showing perfect energy balancing $\Delta_{\text{DE}} \approx
           \Delta_{\text{FE}}$ (left) and imperfect energy balancing
           $\Delta_{\text{DE}} > \Delta_{\text{FE}}$ (right) between the
           ferroelectric and dielectric. Even with perfect energy balancing, the
           ferroelectric eventually runs out of energy stored in its phase
           transition. This occurs at the end of the linear region, after which
           point the ferroelectric can no longer supply energy to the dielectric
           and must receive energy from an external source to continue
           polarizing. (b) Total energy dissipated as a function of the
           capacitance matching $|C_{\text{FE}} / C_{\text{DE}}|$ after storing
           charge $Q$ on the capacitor plates. $C_{\text{FE}}$ is the
           ferroelectric capacitance, and $C_{\text{DE}}$ is the dielectric
           capacitance. The energy dissipated is normalized to $1 / 2 \,
           C_{\text{DE}} V_{\text{DE}}^{2}$, which is the energy conventionally
           dissipated without a ferroelectric. $Q_{\text{crit}}$ is the charge
           corresponding to the end of the linear region. (c) Total energy
           dissipated as a function of the capacitance matching $|C_{\text{FE}}
           / C_{\text{DE}}|$ after discharging $Q$ amount of charge.}
  \label{fig:matching}
\end{figure*}

The capacitance matching can be better understood by examining the energy
landscapes shown in Fig. \ref{fig:matching}a.
In the left set of landscapes, the energy available in the ferroelectric phase
transition closely matches the energy ``needs'' of the dielectric in the linear
region. This allows the ferroelectric to supply nearly all of the energy needed
to charge the dielectric. Consequently, minimal additional energy is needed from
an external source, and less energy will be lost to radiation while propagating
from the source to the dielectric. This is an example of perfect energy
balancing. In contrast, the right half of Fig. \ref{fig:matching}a shows an
example of poor energy balancing. The energy available in the ferroelectric is
insufficient for charging the dielectric. Consequently, more energy is needed
from an external source, and more energy will be lost to radiation while
propagating to the dielectric. We can control the energy balancing by tuning the
energy landscapes. This is accomplished by changing film thickness or electric
susceptibility (e.g. by changing temperature or strain; or using different
materials). Since these parameters directly affect the system's capacity to
store energy in an electric field, the energy balancing can be thought of as a
capacitance matching between the ferroelectric and dielectric. In the linear
region, for example, the ferroelectric capacitance is negative due to the
negative curvature of the energy landscape and should be equal and opposite to
the dielectric capacitance for proper matching. The ferroelectric is able to
supply energy to the dielectric within this linear region. However, it will
eventually run out of stored energy and begin requiring energy from an external
source to continue polarizing. This is reflected by the change in the energy
landscape's curvature from negative to positive at the end of the linear region.
This is also reflected in the charge dependency of the energy dissipation as
shown in Fig. \ref{fig:matching}b-c. No matter how perfectly matched the
ferroelectric and dielectric are, the energy dissipation increases for greater
amounts of charge during both charging (Fig. \ref{fig:matching}b) and
discharging (Fig. \ref{fig:matching}c).

\begin{figure*}
  \includegraphics[width=0.8\textwidth]{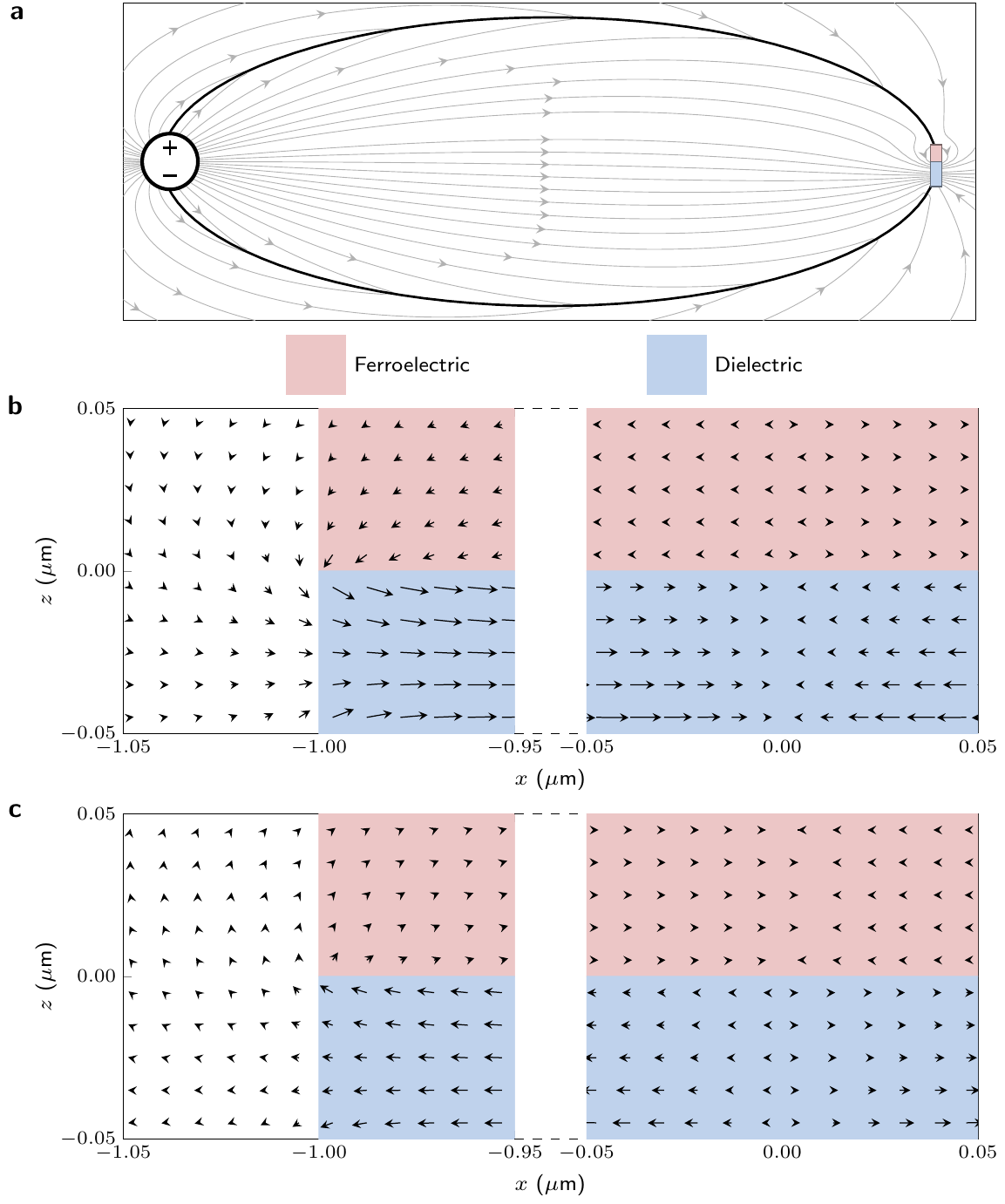}
  \caption{Poynting vector calculations. (a) Schematic of overall energy flow
           from the source to the ferroelectric-dielectric capacitor during
           charging. (b) Poynting vector field along the side of the capacitor
           and near the center during charging. $x$ is the in-plane spatial
           coordinate, and $z$ is the spatial coordinate along the capacitor
           axis. (c) Poynting vector field along the side of the capacitor and
           near the center during discharging.}
  \label{fig:poynting}
\end{figure*}

Finally, we explicitly show how energy transfers between the ferroelectric and
dielectric by computing the Poynting vector at various positions in and around
the system. The overall energy flow during charging is shown schematically in
Fig. \ref{fig:poynting}a.
Notice that energy flows from the source to the ferroelectric-dielectric system,
but some of it radiates away. If we zoom in on the capacitor (Fig.
\ref{fig:poynting}b), we see that energy flows directly from the ferroelectric
to the dielectric through the surrounding space. The Poynting vector diverges
outwards (energy is decreasing) at the center of the ferroelectric and inwards
(energy is increasing) at the dielectric center. During discharging (Fig.
\ref{fig:poynting}c), the Poynting vector diverges oppositely compared to the
charging case. The dielectric acts as the source with some of its energy flowing
into the ferroelectric through the surrounding space while the remainder
radiates away.

In conclusion, we have shown that it is possible to improve upon the otherwise
fundamental limit on energy dissipation of $1 / 2 \, C V^{2}$ during charging
and discharging of a capacitor by coupling a ferroelectric to the dielectric. We
used a thermodynamic model to show that the dielectric can stabilize the
ferroelectric near its phase transition, enabling extraction of the energy
stored in the phase transition. Poynting's theorem and Maxwell's equations then
explicitly showed that this stored energy directly flows between the
ferroelectric and dielectric during charging and discharging. The net result is
a reduction in total energy dissipation below the conventional limit. This
reduction can be maximized by balancing the energy stored in the ferroelectric
phase transition with the energy needed by the dielectric. These results provide
the framework for understanding and improving fundamental energy efficiency in
all devices that operate on storing energy in electric fields.

\appendix

\section{Poynting's Theorem}
Poynting's theorem is a statement of conservation of energy for a system of
charged particles and can be written as a differential continuity equation as
given in (\ref{eq:poynting_theorem}). The integral form is
\begin{equation}
  \int_{\Omega} \left( \frac{\partial u}{\partial t} + \nabla \cdot \mathbf{S}
    \right) \, d^{3}r = -\int_{\Omega} (\mathbf{J} \cdot \mathbf{E}) \, d^{3}r
\end{equation}
where the volume $\Omega$ can be arbitrary, but we take it as the volume filled
by the components of the circuit to establish appropriate boundary conditions.
This differential equation can be numerically solved for the current density
$\mathbf{J}$ if the remaining variables can be expressed in terms of current
density. We accomplish this by establishing a consistent relationship between
internal energy density $u$ and electric field $\mathbf{E}$ for a given current
density. First, an initial electric field is assumed, and the state of the
ferroelectric-dielectric system is determined by minimizing the free energy in
(\ref{eq:free_energy}) with respect to small polarization fluctuations. This
determines the ferroelectric and dielectric polarization states, which allow us
to determine the internal energy density $u$ and charge density $\rho$. This
charge density can be used in conjunction with a given current density to solve
Maxwell's equations by determining the retarded scalar and vector potentials:
\begin{multline} \label{eq:retarded_scalar_potential}
  V(\mathbf{r}, t) = \\\frac{1}{4 \pi \epsilon_{0}} \int
    \frac{\rho(\mathbf{r}', t')}{|\mathbf{r} - \mathbf{r}'|} \delta \left( t' - t +
    \frac{|\mathbf{r} - \mathbf{r}'|}{c} \right) \, dt' \, d^{3}r'
\end{multline}
\begin{multline} \label{eq:retarded_vector_potential}
  \mathbf{A}(\mathbf{r}, t) = \\\frac{\mu_{0}}{4 \pi} \int
    \frac{\mathbf{J}(\mathbf{r}', t')}{|\mathbf{r} - \mathbf{r}'|} \delta
    \left( t' - t + \frac{|\mathbf{r} - \mathbf{r}'|}{c} \right) \, dt' \,
    d^{3}r'
\end{multline}
The solutions to Maxwell's equations are then
\begin{align}
  \mathbf{E}(\mathbf{r}, t) &= -\nabla V - \frac{\partial \mathbf{A}}{\partial
    t} \label{eq:electric_field} \\
  \mathbf{B}(\mathbf{r}, t) &= \nabla \times \mathbf{A}
    \label{eq:magnetic_field}
\end{align}
To reduce numerical error, we directly solved Maxwell's equations by
analytically combining
(\ref{eq:retarded_scalar_potential})-(\ref{eq:magnetic_field}) and evaluating
the resulting Jefimenko's equations. Notice that the electric field computed
here must be equal to the electric field initially assumed when determining
the internal energy density. Thus, the electric field and internal energy
density must be solved for consistently, and this can be accomplished for a
given current density. From (\ref{eq:retarded_vector_potential}) and
(\ref{eq:magnetic_field}), the magnetic field \(\mathbf{B}\) only depends on
current density, so the Poynting vector \(\mathbf{S}\) can also be computed
for a given current density:
\begin{equation}
  \mathbf{S} = \frac{\mathbf{E} \times \mathbf{B}}{\mu_{0}}
\end{equation}
The only remaining unknown quantity is the current density, which can now be
solved for using Poynting's theorem.

\begin{acknowledgments}
  This work was supported by the Berkeley Center for Negative Capacitance
  Transistors. JCW acknowledges generous support from an NSF graduate research
  fellowship.
\end{acknowledgments}

\bibliographystyle{unsrtnat}
\bibliography{References,references-SS}

\end{document}